\documentclass[preprint,12pt]{elsarticle}
\usepackage{amssymb}
\usepackage{amsmath}
\usepackage{physics}
\usepackage{geometry} 
\newgeometry{vmargin={3cm,3.5cm}, hmargin={3.5cm,3.5cm}} 

\journal{Journal of Subatomic Particles and Cosmology}

\begin{document}
\newcommand\beq{\begin{equation}}
\newcommand\eeq[1]{\label{#1}\end{equation}}
\newcommand\eqn[1]{Eq.\ (\ref{#1})}
\newcommand{\md}{m_{\scriptscriptstyle D}}
\newcommand{\mq}{M_{\scriptscriptstyle Q}}
\newcommand{\pq}{p_{\scriptscriptstyle Q}}
\newcommand{\eb}{E_{\scriptscriptstyle B}}
\newcommand{\qqb}{\bar{Q} Q}
\newcommand{\jpsi}{J/\Psi}
\newcommand{\snn}{\sqrt{s}_{\scriptscriptstyle NN}}
\newcommand{\gee}{G_{\scriptscriptstyle EE}}
\newcommand{\geeg}{G_{\scriptscriptstyle EE}^>}
\newcommand{\get}{G^{\scriptscriptstyle E}}
\newcommand{\gert}{G^{\scriptscriptstyle E}_r}
\newcommand{\gef}{G^{\scriptscriptstyle E}_{\scriptscriptstyle F}}
\newcommand{\goct}{G^{\scriptscriptstyle E}_{\scriptscriptstyle oct}}
\newcommand{\tc}{T_c}
\newcommand{\nc}{N_c}
\newcommand{\cf}{C_{\scriptscriptstyle F}}
\def\up(#1){\Upsilon({#1}S)}

\begin{frontmatter}

\title{Lattice study of correlators for quarkonium decay}

\author[label1]{Saumen Datta}
\author[label2]{Debasish Banerjee}
\author[label3]{Nora Brambilla}
\author[label3]{Marc Janer}
\author[label4]{Viljami Leino}
\author[label3]{Julian Mayer-Steudte}
\author[label5]{Peter Petreczky}
\author[label6]{Balbeer Singh}
\author[label3]{Antonio Vairo}

\affiliation[label1]{organization={Tata Institute of Fundamental Research},
            addressline={Homi Bhabha Road}, 
            city={Mumbai},
            postcode={400005}, 
            country={India}}

\affiliation[label2]{organization={School of Physics and Astronomy,
    University of Southampton},
  postcode={SO17 1BJ}, 
  country={UK}}

\affiliation[label3]{organization={Technical University of Munich, TUM
    School of Natural Sciences, Physics Department},
  addressline={James-Franck-Strase 1},
  city={Garching},
  postcode={85748},country={Germany}}

\affiliation[label4]{organization={Institut fuer Kernphysik, Johannes
    Gutenberg-Universitaet Mainz},
  addressline={Johann-Joachim-Becher-Weg 48},
  city={Mainz},postcode={55128},country={Germany}}

\affiliation[label5]{organization={Physics Department, Brookhaven National
    Laboratory},addressline={Upton},city={New York},postcode={11973},
  country={USA}}

\affiliation[label6]{organization={Department of Physics, University of
    South Dakota},city={Vermillion},
            postcode={57069}, 
            state={South Dakota},
            country={USA}}

\begin{abstract} While there has been a lot of progress in developing a
  formalism for the study of quarkonia in QGP, a nonperturbative
  study is still difficult. For bottomonia, where the system size
  is much less than the inverse temperature, the interaction of the
  system with the medium can be approximated by a dipole interaction with
  the color electric field. The decay of the quarkonia can be connected to
  a correlation function of the color electric field.

  We present preliminary results from a lattice study of the relevant color
  electric field correlator. The structure of the correlator, and its
  difference from the corresponding correlator studied for heavy quark
  diffusion, is discussed.
\end{abstract}

\end{frontmatter}

Quarkonia are among the most-studied probes of the medium created
in heavy ion collision experiments. The suppression of quarkonia as a
signal of the formation of a deconfined medium was suggested long back by
Matsui and Satz \cite{satz}. While quarkonia
suppression has been observed in relativistic heavy ion collision
experiments, the theoretical calculation of the suppression is
involved.

\begin{figure}[t]
\centering
\includegraphics[width=7.5cm,height=3.6cm]{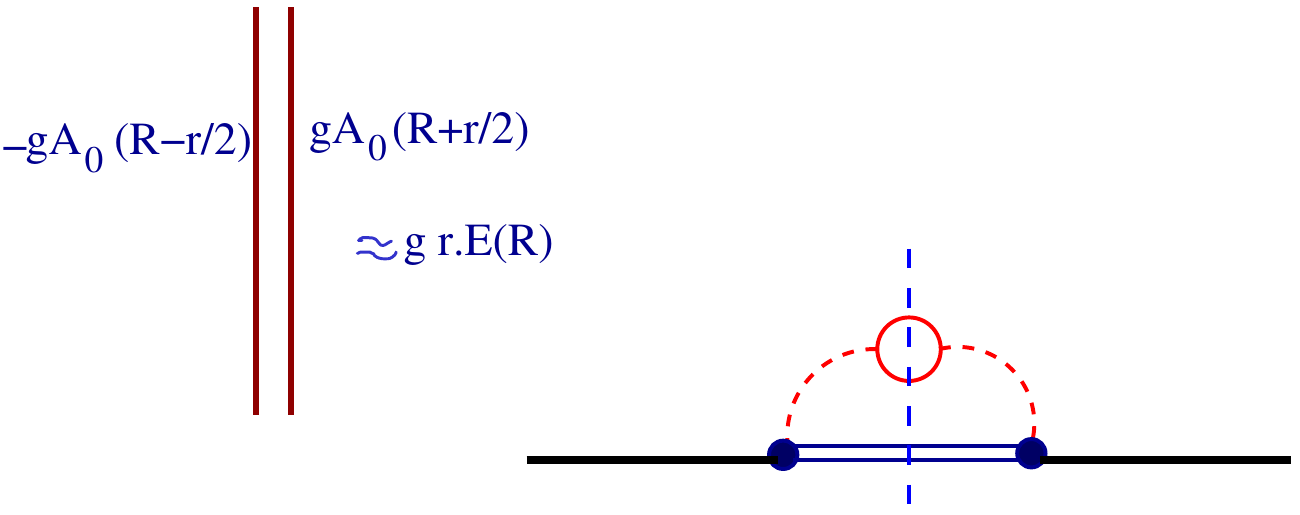}\hskip1.0cm\includegraphics[width=5.6cm,height=3.6cm]{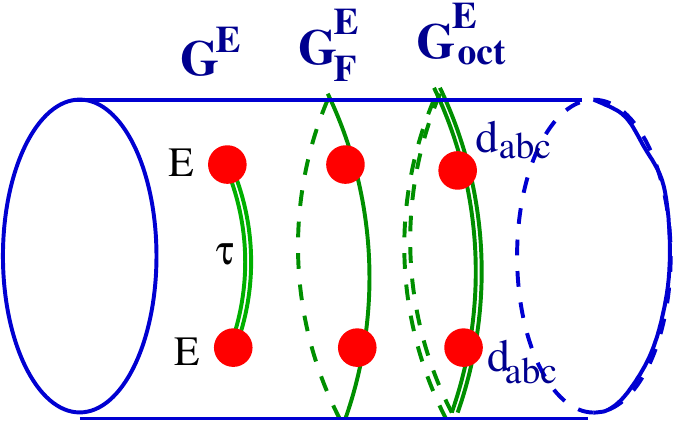}
\caption{(Left) The interaction of bottomonia with thermal gluons for
  $\qqb$ separation $r \ll 1/T$, where $T$ is the temperature of the medium.
  (Middle) Bottomonia decay in $pNRQCD$ due to singlet $\to$ octet transition.
  (Right) Various EE correlators: $\get(\tau)$ (\eqn{get}), $\gef(\tau)$
  (\eqn{gfund}), $\goct(\tau)$ (\eqn{goct}). The single and double lines
  connecting the $E$ fields denote fundamental and adjoint Wilson lines,
  respectively.}
\label{fig1}
\end{figure}

The scales
of interest of the thermal medium are the temperature $T$ and the Debye
mass $\md$.
The heavy quark mass $\mq$ can be safely taken to be larger than these
scales. The comparison of the medium scales with the inverse size of
the quarkonia $1/r \sim \pq=\mq v$ and the binding energy scale
$\eb \approx \mq v^2$ determine the theoretical treatment \cite{pNRQCD-T}.
For ground state bottomonia, even at LHC energies
$1/r \gg T$. Then the thermal gluons cannot resolve the $Q, \bar{Q}$, and
the interaction of the medium with the quarkonia is like a color dipole
$g \, \vec{r}.\vec{E}$ (Fig. \ref{fig1}).

The proper framework for a theoretical study of the modification of
quarkonia yield due to the deconfined medium is that of the open
quantum system\cite{akamatsu}: the $\qqb$+environment (thermal medium)
density matrix is evolved in time, and then the environment
is traced out to get the density matrix for the $\qqb$ system. For bottomonia,
the formalism can be written in a tractable form by using the potential
nonrelativistic QCD (pNRQCD) formalism \cite{pnrqcd}. In this formalism, the
degrees of freedom are the $\qqb$ in singlet $\vert s \rangle$ and octet
$\vert o \rangle$ at different separations $r$. The quarkonia decay is
given by a singlet $\to$ octet transition (Fig. \ref{fig1}). The decay
width can be written as
\beq
\Gamma \; = \; \int d^3p \, \lvert \langle \psi_b \vert \vec{r} \vert
\psi_{\vec p} \rangle \rvert^2 \ \geeg (\Delta E), \qquad
  \geeg(t) \; = \; \frac{g^2 T_F}{3 N_c} \; \langle E_i^a(t)
  \, W^{ab}(t, 0) \, E_i^b(0) \rangle_T
\eeq{gee}
where $\Delta E$ is the singlet-octet energy difference,
$\langle \psi_b \lvert \vec{r} \rvert \psi_{\vec p} \rangle$
is the amplitude for the singlet to octet transition, $i$ denotes
the space index, $ W^{ab}(t, 0)$ is the adjoint Wilson line
connecting the points $(0, \vec{x})$ and $(t, \vec{x})$, and an averaging
over the spatial coordinate $\vec{x}$ is implied. The medium
effect is completely encoded in the color electric correlator $\geeg$. 

For $T \gg \eb$, the binding energy of the quarkonium, one can use the 
Markovian description, and write a Lindblad equation
for the density matrix \cite{akamatsu}, in which the medium
effect is completely
encoded in a few transport coefficients \cite{tum1}, the most important one
being $\kappa$ obtained from $\geeg(\Delta E \to 0)$ in \eqn{gee}. This
formalism has been used to calculate the yield of the $\Upsilon$ states
\cite{tum2}, where the transport coefficients were taken from a three-loop
calculation. Also the non-Markovian correction has been estimated
\cite{vyshakh}. A classical Boltzmann description has also been written
down \cite{mehen}. 

$\geeg(t)$ can be continued to the Euclidean time correlator \cite{yao}
\beq
\get(\tau) \; = \; - \frac{1}{3} \sum_i \langle E_i^a(\tau) \, W^{ab}(\tau, 0)
\, E_i^b(0) \rangle_T \, .
\eeq{get}
where $\tau$ denotes Euclidean time. It has been calculated 
perturbatively at NLO, both in real-time \cite{yao}
and in the imaginary time formalism \cite{tumpert}, and has recently
been calculated nonperturbatively using lattice QCD \cite{tumlat}.
In this report we present lattice results for $\get(\tau)$ at various
temperatures in a gluonic plasma.

$\get(\tau;a)$ has a mass divergence coming from 
$W^{ab}$, which needs to be renormalized before taking the lattice spacing
$a \to 0$. It is illuminating to compare \eqn{get} with the correlator
\beq
\gef(\tau) \; = \; -\frac{1}{3 L_f} \sum_i \langle \Re \tr W_f(1/T, \tau) \,
E_i(\tau) \, W_f(\tau, 0) \, E_i(0) \rangle_T
\eeq{gfund}
used for the study of heavy quark diffusion in quark-gluon plasma
\cite{clm}. $\gef(\tau)$ has been calculated on the lattice for both
gluonic \cite{hqdrnf0} and 2+1 flavor QCD \cite{hqdrnf3}.
Here $W_f(\tau_1, \tau_2)$ is a fundamental Wilson line, and $L_f=\frac{1}{3}
\tr W_f(1/T, 0)$ is the fundamental Polyakov loop. \eqn{gfund} differs from
\eqn{get} in the orientation of the Wilson line. The mass
divergence of $W_f$ gets cancelled by the corresponding one from $L_f$.
Motivated by this, we look at the renormalized correlator \cite{tumlat}
\beq
\gert(\tau) \; = \; e^{\delta m(a) \, \tau} \get(\tau; a) \; = \;
\left(\frac{L_a^r(T)}{L_a(a; T)} \right)^{\tau T} \get(\tau; a)
\eeq{gert}
where $L_a^r(T)$ and $L_a(a;T)$ are the renormalized adjoint Polyakov
loop and the adjoint Polyakov loop, respectively, measured on a lattice
of spacing $a$.

\begin{figure}
  \centering
  \includegraphics[width=7.0cm]{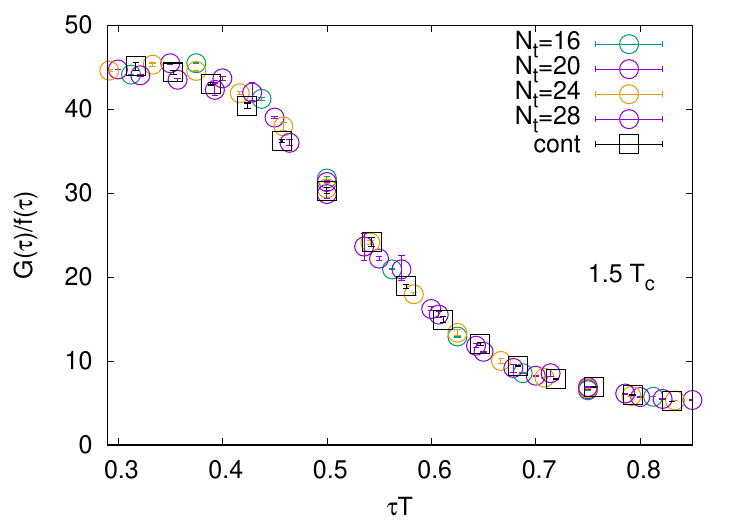}\includegraphics[width=7.0cm]{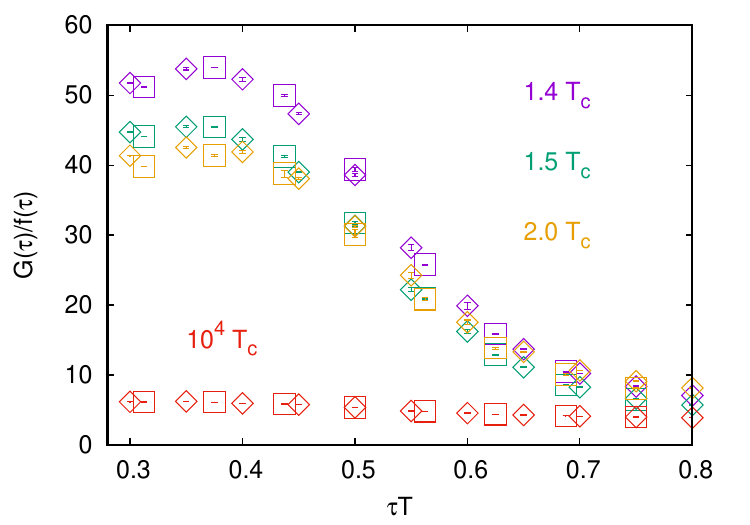}
  \caption{(left) $\gert(\tau;a)$ at different lattice spacings and the
    continuum extrapolation, at 1.5 $\tc$. (Right) $\gert(\tau;a)/f(\tau)$
    at different temperatures. At each temperature results for two different
    lattice spacings are shown.}
  \label{cutoff}
\end{figure}

The major part of the cutoff effect in $\get(\tau, a)$ comes from the
exponential mass divergence, which is taken care of by \eqn{gert}.
In the left panel of Fig. \ref{cutoff} we show the results of a lattice
calculation of
$\gert(\tau;a)$ at 1.5 $\tc$ at different lattice spacings $a=1/N_\tau \, T$.
In order to better illustrate the structure of the correlator, we have
normalized it with 
\beq
f(\tau;a) \; = \; \frac{\gert(\tau;a)\vert_{\scriptscriptstyle LO}}{g^2 \,
  (\nc^2-1)}
\eeq{gnorm}
where the leading order calculation is done for the lattice discretized
correlator. In the range $\tau T \in [0.3,0.8]$ we find mild cutoff
effect in the renormalized correlator. The continuum extrapolated correlator
is also shown in Fig. \ref{cutoff}.

The remarkable feature of the nonperturbative correlator in Fig.
\ref{cutoff} is that it is not symmetric about $\tau T=1/2$. This is in
contrast to both the leading order behavior,
$\gef(\tau)\vert_{\scriptscriptstyle LO}$, and the correlator $\gef(\tau)$.
Note that at leading order, the Wilson line does not contribute,
and therefore $\gef(\tau)$ differs from the adjoint correlator only by a color
factor: $\gef(\tau, a)\vert_{\scriptscriptstyle LO} \; = \; g^2 \, \cf \,
f(\tau; a)$ where $\cf=\tfrac{\nc^2-1}{2 \nc}$. 
The difference in symmetry properties of $\get(\tau)$ and $\gef(\tau)$
originate from the different orientations of the Wilson lines, and can be
understood from Fig. \ref{fig1}. The
antisymmetric part can also be seen in the NLO correlator \cite{yao,tumpert},
where it appears at $\mathcal{O}(g^4)$. Therefore, at high temperatures, where
$g$ is small, the antisymmetric contribution is expected to decrease.
This can be seen in the right panel of Fig. \ref{cutoff}, where $\gert(\tau)$
is shown at different temperatures.

\begin{figure}
  \centering
  \includegraphics[width=7.0cm]{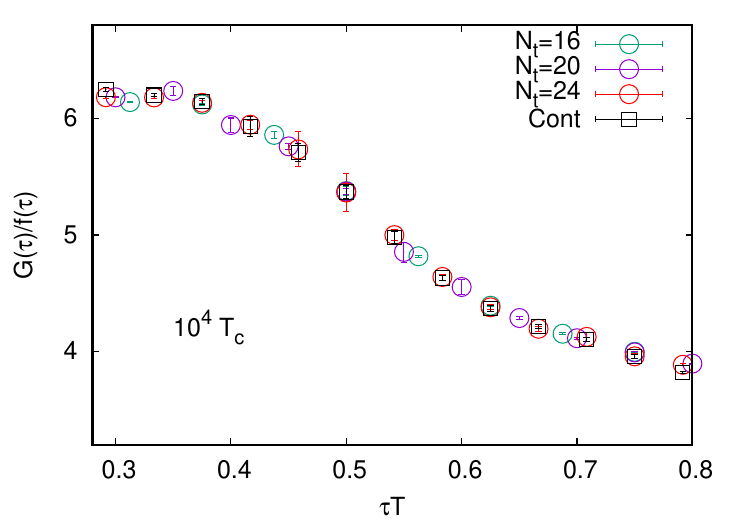}
  \includegraphics[width=7.0cm]{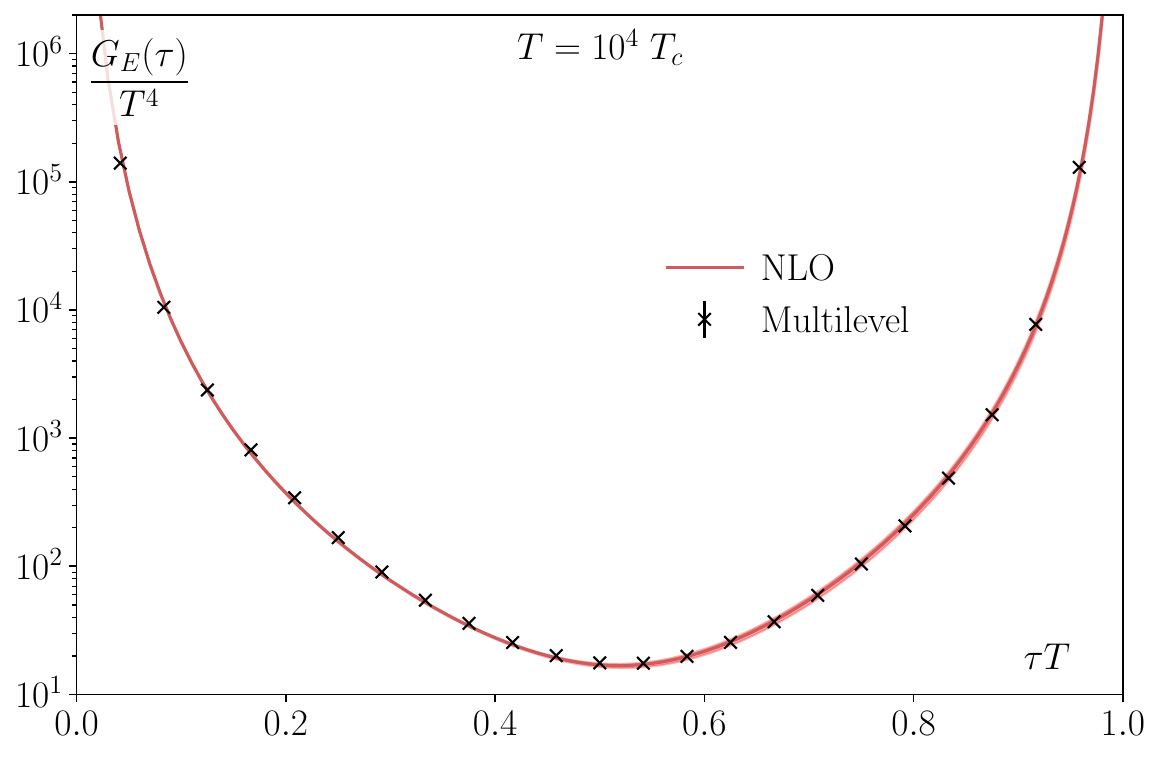}
  \caption{$\gert(\tau)$ at $10^4 T_c$. (Left) Results at different lattice
    spacings, and the continuum limit. 
    (Right) Comparison of the continuum $\gert(\tau; T=10^4 \tc)$
    with the NLO result of
    Ref. \cite{tumpert}.}
  \label{pert}
\end{figure}

At very high temperatures, we expect the lattice correlator to approach the
perturbative result. In Fig. \ref{pert} we show the correlator calculated at
$10^4 T_c$, and its comparison with the NLO result \cite{tumpert}. The
nonperturbative result agrees very well with the NLO result at this
high temperature.

For quarkonia phenomenology, the interesting quantity is $\geeg({\rm low} \
\omega)$. In particular, in the Markovian description \cite{tum1} we are
interested in the transport coefficient $\kappa$ obtained from the $\omega
\to 0$ limit. The calculation of this is ongoing.

Here, we have concentrated on the correlator $\get(\tau)$, which is the
most important one in the study of quarkonia in QGP \cite{tum1,mehen}.
In the pNRQCD description of quarkonia in QGP, one also needs the
octet-octet interaction term, which leads to the correlator 
\beq
\goct(\tau) \; = \; -\frac{1}{3 L_a} \sum_i \langle W^{ab}(1/T, \tau) \,
d_{bcd} \, E_i^d(\tau) \, W^{ce}(\tau, 0) \, d_{eaf} \, E_i^f(0) \rangle_T \, .
\eeq{goct}
$d_{abc}$, the symmetric
structure functions of SU(3), come from the $O^\dagger \left\{ g \, \vec{r}.
\vec{E} \, ,
\, O \right\}$ interaction term in pNRQCD. This correlator has been studied on
the lattice \cite{tumlat}. It is symmetric around $t T=1/2$ (Fig. \ref{fig1}).
Within errors, the nonperturbative correlator obeys the NLO scaling
\[ \tfrac{\goct(\tau;T)}{\gef(\tau;T)} \, \approx \, 2
\tfrac{\nc^2-4}{\nc^2-1} \, . \]


\begin{thebibliography}{00}

\bibitem{satz}
  T. Matsui and H. Satz, 
  Phys. Lett. B 178 (1986) 416.
\bibitem{pNRQCD-T}
  N. Brambilla, J. Ghiglieri, P. Petreczky and A. Vairo,
  Phys. Rev. D78 (2008) 014017.
\bibitem{akamatsu}
  Y. Akamatsu, Phys. Rev. D 87 (2013) 045016; Phys. Rev. D 91 (2015) 056002.
\bibitem{pnrqcd}
  A. Pineda and J. Soto, Nucl. Phys. B Proc. Suppl. 64 (1998) 428. \\
  N. Brambilla, A. Pineda, J. Soto and A. Vairo, Nucl. Phys. B 566 (2000) 275.
\bibitem{tum1}
  N. Brambilla, M. Escobedo, J. Soto and A. Vairo, Phys. Rev. D 96 (2017)
  034201; Phys. Rev. D97 (2018) 074009.
\bibitem{tum2}
  N. Brambilla, T. Magorsch, M. Strickland, A. Vairo and P.V. Griend,
  Phys. Rev. D 109 (2024) 114016.
\bibitem{vyshakh}
  Vyshakh B.R. and Rishi Sharma, arXiv:2504.19348.
\bibitem{mehen}
  X. Yao and T. Mehen, JHEP 02 (2021) 062.
\bibitem{yao}
  B. Scheihing-Hitschfeld and X. Yao, Phys. Rev. D 108 (2023) 054024.
\bibitem{tumpert}
  N. Brambilla, P. Panayiotou, S. S\"appi and A. Vairo,
  arXiv:2505.16604.
\bibitem{tumlat}
  N. Brambilla, S. Datta, M. Janer, V. Leino, J. Mayer-Steudte,
  P. Petreczky and A. Vairo, arXiv:2505.16603.
\bibitem{clm}
  S. Caron-Huot, M. Laine and G. Moore, JHEP 04 (2009) 053.
\bibitem{hqdrnf0}
  D. Banerjee, S. Datta, R. Gavai and P. Majumdar, Phys. Rev. D 85 (2012)
  014510; Nucl. Phys. A 1038 (2023) 122721. \\
  L. Altenkort, A.M. Eller, O. Kaczmarek, L. Mazur, G.D. Moore and H-T.Shu,
  Phys. Rev. D 103 (2021) 014511. \\
  N. Brambilla, V. Leino, J. Mayer-Steudte and P. Petreczky, Phys. Rev. D 107
  (2023) 054508.
\bibitem{hqdrnf3}
  L. Altenkort, O. Kaczmarek, R. Larsen, S. Mukherjee, P. Petreczky, H-T. Shu
  and S. Stendebach, Phys. Rev. Lett. 130 (2023) 231902.
\end{thebibliography}
\end{document}